\documentstyle[twoside,epsf,amssymb]{article}
\catcode`\@=11
\long\def\@makefntext#1{
\protect\noindent \hbox to 3.2pt {\hskip-.9pt  
$^{{\eightrm\@thefnmark}}$\hfil}#1\hfill}       

\def\@makefnmark{\hbox to 0pt{$^{\@thefnmark}$\hss}}    
    
\def\ps@myheadings{\let\@mkboth\@gobbletwo
\def\@oddhead{\hbox{}
\rightmark\hfil\eightrm\thepage}   
\def\@oddfoot{}\def\@evenhead{\eightrm\thepage\hfil
\leftmark\hbox{}}\def\@evenfoot{}
\def\sectionmark##1{}\def\subsectionmark##1{}}

\oddsidemargin=\evensidemargin
\newcounter{sectionc}\newcounter{subsectionc}\newcounter{subsubsectionc}
\renewcommand{\section}[1] {\vspace{12pt}\addtocounter{sectionc}{1} 
\setcounter{subsectionc}{0}\setcounter{subsubsectionc}{0}\noindent 
    {\tenbf\thesectionc. #1}\par\vspace{5pt}}
\renewcommand{\subsection}[1] {\vspace{12pt}\addtocounter{subsectionc}{1} 
\setcounter{subsubsectionc}{0}\noindent 
{\bf\thesectionc.\thesubsectionc. {\kern1pt \bfit #1}}\par\vspace{5pt}}
\renewcommand{\subsubsection}[1] {\vspace{12pt}\addtocounter{subsubsectionc}{1}
    \noindent{\tenrm\thesectionc.\thesubsectionc.\thesubsubsectionc.
    {\kern1pt \tenit #1}}\par\vspace{5pt}}
\newcommand{\nonumsection}[1] {\vspace{12pt}\noindent{\tenbf #1}
    \par\vspace{5pt}}

\newcounter{appendixc}
\newcounter{subappendixc}[appendixc]
\newcounter{subsubappendixc}[subappendixc]
\renewcommand{\thesubappendixc}{\Alph{appendixc}.\arabic{subappendixc}}
\renewcommand{\thesubsubappendixc}
    {\Alph{appendixc}.\arabic{subappendixc}.\arabic{subsubappendixc}}

\renewcommand{\appendix}[1] {\vspace{12pt}
        \refstepcounter{appendixc}
        \setcounter{figure}{0}
        \setcounter{table}{0}
        \setcounter{lemma}{0}
        \setcounter{theorem}{0}
        \setcounter{corollary}{0}
        \setcounter{definition}{0}
        \setcounter{equation}{0}
        \renewcommand{\thefigure}{\Alph{appendixc}.\arabic{figure}}
        \renewcommand{\thetable}{\Alph{appendixc}.\arabic{table}}
        \renewcommand{\theappendixc}{\Alph{appendixc}}
        \renewcommand{\thelemma}{\Alph{appendixc}.\arabic{lemma}}
        \renewcommand{\thetheorem}{\Alph{appendixc}.\arabic{theorem}}
        \renewcommand{\thedefinition}{\Alph{appendixc}.\arabic{definition}}
        \renewcommand{\thecorollary}{\Alph{appendixc}.\arabic{corollary}}
        \renewcommand{\theequation}{\Alph{appendixc}.\arabic{equation}}
        \noindent{\tenbf Appendix \theappendixc #1}\par\vspace{5pt}}
\newcommand{\subappendix}[1] {\vspace{12pt}
        \refstepcounter{subappendixc}
        \noindent{\bf Appendix \thesubappendixc. {\kern1pt \bfit #1}}
    \par\vspace{5pt}}
\newcommand{\subsubappendix}[1] {\vspace{12pt}
        \refstepcounter{subsubappendixc}
        \noindent{\rm Appendix \thesubsubappendixc. {\kern1pt \tenit #1}}
    \par\vspace{5pt}}

\newcommand{\textlineskip}{\baselineskip=13pt}
\newcommand{\smalllineskip}{\baselineskip=10pt}

\newcommand{\copyrightheading}[1]
    {}

\newcommand{\pub}[1]{}

\newcommand{\publisher}[2]{}

\def\abstracts#1#2#3{{
    \centering{\begin{minipage}{4.5in}\footnotesize\baselineskip=10pt
    \parindent=0pt #1\par 
    \parindent=15pt #2\par
    \parindent=15pt #3
    \end{minipage}}\par}} 

\def\keywords#1{}
\def\communicate#1{}

\renewenvironment{thebibliography}[1]
        {\frenchspacing
     \ninerm\baselineskip=11pt
         \begin{list}{\arabic{enumi}.}
        {\usecounter{enumi}\setlength{\parsep}{0pt}     
     \setlength{\leftmargin 12.7pt}{\rightmargin 0pt}
         \setlength{\itemsep}{0pt} \settowidth
    {\labelwidth}{#1.}\sloppy}}{\end{list}}

\newcounter{itemlistc}
\newcounter{romanlistc}
\newcounter{alphlistc}
\newcounter{arabiclistc}

\newcommand{\fcaption}[1]{
        \refstepcounter{figure}
        \setbox\@tempboxa = \hbox{\footnotesize Fig.~\thefigure. #1}
        \ifdim \wd\@tempboxa > 5in
           {\begin{center}
        \parbox{5in}{\footnotesize\smalllineskip Fig.~\thefigure. #1}
            \end{center}}
        \else
             {\begin{center}
             {\footnotesize Fig.~\thefigure. #1}
              \end{center}}
        \fi}

\newcommand{\tcaption}[1]{
        \refstepcounter{table}
        \setbox\@tempboxa = \hbox{\footnotesize Table~\thetable. #1}
        \ifdim \wd\@tempboxa > 5in
           {\begin{center}
        \parbox{5in}{\footnotesize\smalllineskip Table~\thetable. #1}
            \end{center}}
        \else
             {\begin{center}
             {\footnotesize Table~\thetable. #1}
              \end{center}}
        \fi}

\def\pmb#1{\setbox0=\hbox{#1}
    \kern-.025em\copy0\kern-\wd0
    \kern.05em\copy0\kern-\wd0
    \kern-.025em\raise.0433em\box0}

\def\fnm#1{$^{\mbox{\scriptsize #1}}$}
\def\fnt#1#2{\footnotetext{\kern-.3em
    {$^{\mbox{\scriptsize #1}}$}{#2}}}

\def\fpage#1{\begingroup
\voffset=.3in
\thispagestyle{empty}\begin{table}[b]\centerline{\footnotesize #1}
    \end{table}\endgroup}

\def\runninghead#1#2{\pagestyle{myheadings}
\markboth{{\protect\footnotesize\it{\quad #1}}\hfill}
{\hfill{\protect\footnotesize\it{#2\quad}}}}
\font\tenrm=cmr10
\font\tenit=cmti10 
\font\tenbf=cmbx10
\font\bfit=cmbxti10 at 10pt
\font\ninerm=cmr9

\font\eightrm=cmr8

\def\FigName{figure}%
\newbox\captionbox
\long\def\@makecaption#1#2{%
  \ifx\FigName\@captype
    \vskip\abovecaptionskip
    \setbox\tempbox\hbox{{\figurecaptionfont #1\hskip1em #2}}
    \ifdim\wd\tempbox< 28pc
    \centerline{\box\tempbox}
    \else
    {\figurecaptionfont #1\hskip1em #2\par}
\fi\else
    \setbox\tempbox\hbox{{\tablecaptionfont #1\hskip1em #2}}
    \ifdim\wd\tempbox< 28pc 
    \centerline{\box\tempbox}
    \else
    {\tablecaptionfont #1\hskip1em #2\par}%
    \fi   
 \vskip\belowcaptionskip
 \fi}
\InputIfFileExists{psfig.sty}
{\typeout{^^Jpsfig.sty inputed...ok}}{\typeout{^^JWarning: psfig.sty could be be found.^^J}}
\InputIfFileExists{epsfsafe.tex}
{\typeout{^^Jepsfsafe.tex inputed...ok}}
            {\typeout{^^JWarning: epsfsafe.tex could not be found.^^J}}
\InputIfFileExists{epsfig.sty}
{\typeout{^^Jepsfig.sty inputed...ok}}{\typeout{^^JWarning: epsfig.sty could not be found.^^J}}
\InputIfFileExists{epsf.sty}
{\typeout{^^Jepsf.sty inputed...ok}}{\typeout{^^JWarning: epsf.sty could not be found.^^J}}%
\def\fps@figure{tbp}
\def\ftype@figure{1}
\def\ext@figure{lof}
\def\fnum@figure{Fig.\ \thefigure}
\def\qed{\hbox{${\vcenter{\vbox{              
   \hrule height 0.4pt\hbox{\vrule width 0.4pt height 6pt
   \kern5pt\vrule width 0.4pt}\hrule height 0.4pt}}}$}}

\newcommand{\be}{\begin{equation}}
\newcommand{\bea}{\begin{eqnarray}}
\newcommand{\eea}{\end{eqnarray}}
\newcommand{\ee}{\end{equation}}

\newcommand{\bra}[1]{\mbox{$\langle #1 |$}}
\newcommand{\ket}[1]{\mbox{$| #1 \rangle$}}

\newcommand{\proj}[1]{\ket{#1}\!\bra{#1}}

\begin{document}
\setlength{\textheight}{8.0truein}    

\runninghead{Purification of two-qubit mixed states} 
            {Enric Jan\'e}

\normalsize\textlineskip
\thispagestyle{empty}
\setcounter{page}{1}

\copyrightheading{} 

\vspace*{0.88truein}

\fpage{1}
\centerline{\bf
PURIFICATION OF TWO-QUBIT MIXED STATES}
\vspace*{0.37truein}
\centerline{\footnotesize 
ENRIC JAN\'E}
\vspace*{0.015truein}
\centerline{\footnotesize\it Departament d'Estructura i Constituents de la Mat\`{e}ria, Universitat de Barcelona, Diagonal 647}
\baselineskip=10pt
\centerline{\footnotesize\it Barcelona, E-08028, Spain}
\vspace*{0.225truein}
\publisher{(received date)}{(revised date)}

\vspace*{0.21truein}
\abstracts{
We find the necessary and sufficient condition under which two two-qubit mixed states can be purified into a pure maximally entangled state by local operations and classical communication.  The optimal protocol for such transformation is obtained. This result leads to a necessary and sufficient condition for the exact purification of $n$ copies of a two-qubit state.}{}{}

\vspace*{10pt}
\vspace*{3pt}
\communicate{to be filled by the Editorial}

\vspace*{1pt}\textlineskip  
\noindent
Quantum entanglement is regarded as one of the basic resources in the field of quantum information theory. It is the essential ingredient for tasks as quantum dense coding \cite{bennett92} and quantum teleportation \cite{bennett93}. In general these tasks require maximally entangled states, among which the Bell state
\be
\ket{\Psi_-}=\frac{1}{\sqrt{2}}(\ket{01}-\ket{10})
\ee
is a frequent choice. However, in practical situations decoherence makes pure states evolve into mixed states and reduces its entanglement. Therefore, in the last years much work has been devoted to understand under which conditions the entanglement that has been lost can be recovered using local operations and classical communication (LOCC). 

In general there are two frameworks under which the problem of concentrating the entanglement is studied. On the one hand, there are cases in which a large amount of copies of a given state are available. That is, two parties, $A$ and $B$, are given a very large number of copies, $n$, of a bipartite state, $\rho_{AB}^{\otimes n }$, and they have to make operations on their subsystems in order to concentrate the entanglement so that they end up with $m$ copies of the state $\ket{\Psi_-}$. In general the problem is stated in the asymptotic limit ($n\rightarrow\infty$), and the final state after the local operations has a fidelity with the state $\proj{\Psi_-}^{\otimes m}$ going to 1 as $n$ goes to infinity, keeping $\frac{m}{n}$ finite. If the process of concentrating the entanglement is optimal, the ratio $\frac{m}{n}$ is called the Entanglement of Distillation \cite{bennett96} of $\rho_{AB}$.

On the other hand, sometimes only one or a few copies of the state $\rho_{AB}$ are given to the parties. In this case there are different strategies one may follow. For instance, if we are given a state of two qubits we can consider the LOCC protocol that achieves the state maximizing the fidelity with the state $\ket{\Psi_-}$ with a finite probability \cite{verstraete02,kent98,verstraete01}. This procedure ensures the state has the maximal probability of behaving as a maximally entangled state. An entanglement measure as the Entanglement of Formation \cite{bennett96} can be used instead of the fidelity and we can study the LOCC protocol that yields a state maximizing this quantity \cite{kent99}.

Another possible strategy would be to see whether a mixed state can be transformed (purified) into a pure maximally entangled state \cite{chen01,chen01a}. This guarantees that, if the protocol is successful, the final state is maximally entangled. In this paper we study this problem for the case of two-qubit states and give the optimal protocol to purify two two-qubit mixed states. Moreover we find the necessary and sufficient condition to ensure that a finite number of copies $n$ of a mixed state of two qubits can be purified.

We will begin by giving a necessary condition that any mixed state that can be purified has to fulfill. This will give us the basic tools to study the particular case we are interested in, namely that in which we deal with states of two qubits.


Consider a mixed state shared by two parties, $A$ and $B$, and let $n_A$ and $n_B$ be the dimensions of their Hilbert spaces, $H_A$ and $H_B$. The general form of such a state is
\be\label{state1}
\rho_{AB}=\sum_{i=1}^{N}p_i\proj{\psi_i},
\ee
where $\ket{\psi_i}\in H_A\otimes H_B$, $p_i>0$ and $\sum_i p_i=1$. The number of terms in the sum has to  be at least equal to the rank of $\rho_{AB}$, that is $N\geq r(\rho_{AB})$, and the states that appear in (\ref{state1}), $\{\ket{\psi_i}\}_{i=1}^N$, span the range of $\rho_{AB}$, $R(\rho_{AB})$. There are infinitely many different ways to write $\rho_{AB}$ as a sum of projectors (i.e. different realizations), but the above constraints always have to be fulfilled.

We would like to know in which cases this mixed state $\rho_{AB}$ can be purified to a pure entangled state with some finite probability using local operations (i.e. operations made on $H_A$ and $H_B$ separately) and classical communication (LOCC). This is equivalent to requiring that the final state is a maximally entangled state, since we know that we can always concentrate the entanglement of a pure state probabilistically by local operations. Our aim is to find a local transformation such that from the mixed state $\rho_{AB}$ we end up with a maximally entangled state,
\be
\rho_{AB}\rightarrow \proj{\Psi},\hspace{10pt}\ket{\Psi}=\frac{1}{\sqrt{k}}\sum_{i=1}^{k}\ket{e_i}_A \ket{f_i}_B,
\ee
where $\{\ket{e_i}_A\}$($\{\ket{f_i}_B\}$) are orthogonal states in $H_A(H_B)$ and $k\leq d$.

Let us suppose there is a protocol that purifies the state $\rho_{AB}$ into a pure entangled state. This means that there is a chain of (probably correlated) local operations after which, with some finite probability, the initial state $\rho_{AB}$ is transformed into a pure entangled state $\ket{\Psi}$. The whole process can be written as two operators $M_A$ and $N_B$ ($M_AM_A^\dagger,N_BN_B^\dagger\leq1$) acting respectively on $H_A$ and $H_B$ such that
\be\label{operation}
M_A\otimes N_B \rho_{AB} M_A^\dagger\otimes N_B^\dagger =p
\proj{\Psi},
\ee
where $p$ is the probability of success of such process. It is important to note that (\ref{operation}) is not the most general operation, as that would include those operations in which some of the information of the outcome is not kept. However, in our case we want to end up with a pure state, and therefore we have to keep track of the outcome of the whole operation. Since this operation is linear and the r.h.s. of (\ref{operation}) is a pure state, for each projector $\proj{\psi_i}$ appearing in the realization of $\rho_{AB}$ (\ref{state1}) we have either
\be\label{cond1}
M_A\otimes N_B \proj{\psi_i} M_A^\dagger\otimes N_B^\dagger
\propto \proj{\Psi},
\ee
or
\be\label{cond2}
M_A\otimes N_B \proj{\psi_i} M_A^\dagger\otimes N_B^\dagger =0.
\ee

Owing to the fact that the r.h.s of (\ref{cond1}) is a pure state, we can summarize both (\ref{cond1}) and (\ref{cond2}) as
\be\label{cond3}
M_A\otimes N_B \ket{\psi_i} =\sqrt{q_i} \ket{\Psi},
\ee
where the set of coefficients $\{ q_i\}$ contains at least one non-zero element. Notice that (\ref{cond3}) is a necessary and sufficient condition an operator $M_A\otimes N_B$ must satisfy to purify the state $\rho_{AB}$ to a pure entangled state $\ket{\Psi}$. 

From (\ref{cond3}) it follows that the operator $M_A\otimes N_B$ purifies any mixed state $\tilde\rho_{AB}$ with the same range as $\rho_{AB}$ (i.e. $R(\rho_{AB})\equiv R(\tilde\rho_{AB})$). The reason is that any state $\ket{\tilde\psi_i}$ appearing in a realization of $\tilde\rho_{AB}$,
\be
\tilde\rho_{AB}=\sum_{i=1}^{M}\tilde p_i \proj{\tilde\psi_i},
\ee
will be a linear combination of the states $\{\ket{\psi_i}\}$ that span $R(\rho_{AB})$ (and therefore also span $R(\tilde\rho_{AB})$), and we can apply the necessary and sufficient condition (\ref{cond3}),
\bea\label{cond4}
M_A\otimes N_B \ket{\tilde\psi_i} &=&M_A\otimes N_B \sum_j a_j^i\ket{\psi_j}\nonumber\\
&=&\sqrt{\tilde q_i}\ket{\Psi},
\eea
where $\sqrt{\tilde q_i}=|\sum_j a_j^i\sqrt{q_j}|$, which is also satisfied. The fact that at least one element in $\{{\tilde q_i}\}$ is non-zero is because if that was not the case, $R(\rho_{AB})$ would belong to the kernel of $M_A\otimes N_B$, but this cannot be true because our assumption is that $M_A\otimes N_B$ purifies the state $\rho_{AB}$. This result leads us to the necessary condition for the purification of a state $\rho_{AB}$ to be possible.

\vspace*{12pt}
\noindent
{\bf Necessary condition:} A mixed state $\rho_{AB}$ cannot be purified to a pure entangled state if its range can be spanned by product states.

\vspace*{12pt}
\noindent
{\bf Proof:} As we have showed above, a purification protocol that purifies a state $\rho_{AB}$ will also purify any state $\tilde\rho_{AB}$ with the same range as $\rho_{AB}$. In addition, given a space that can be spanned by product states, there is always a separable state with a range being that space. Therefore, if a protocol would purify a state with a range spanned by product states, it would also purify a separable state with the same range, which is impossible as it would transform a separable state into an entangled one. This completes the proof.

This necessary condition follows from \cite{chen01}, where they consider the case where an entangled ancillary system is used to purify a mixed entangled state.

In the case of qubits ($n_A=n_B=2$) it was shown that a single copy of a mixed state cannot be purified \cite{kent98}. The reason is that the local POVMs, $M_A$ and $N_B$, on both subsystem need to have rank 2 (full rank in this case), since otherwise they would destroy the entanglement between the parties. But the operator $M_A\otimes N_B$ in (\ref{operation}) cannot have full rank because it has to map different states into the same one (or project them away). This makes the task of purifying impossible since in the case of two qubits the whole space has dimension 4.

However, when two states of two qubits are available it could happen that they could be purified. That is, if party $A$ holds one subsystem of each of the two states and party $B$ holds the remaining subsystems, they could find a way of sharing pure state entanglement after performing some correlated local operations on their subsystems. Since we know that a mixed state of two qubits cannot be purified, the necessary condition stated above has to be satisfied by each of the two states. This is because if one of the two states had a range spanned by product states it would mean that the purification protocol should work for the other state together with a separable state, which we know is not possible in the case of two qubits.

We will now study the case of two states of two qubits to see in which cases we can purify entangled pure states.

From the necessary condition it is clear that the first step is to know in which cases a subspace of ${\mathbb{C}}^2\otimes {\mathbb{C}}^2$ cannot be spanned by product states. The only interesting cases are when the dimension of the subspace is 2 or 3, since the case of dimension equal to 1 is the case when we have a pure state from the very beginning and in the case when the dimension is 4 the space is of course spanned by product states (it is the whole space ${\mathbb{C}}^2\otimes {\mathbb{C}}^2$).

When the dimension of the subspace of ${\mathbb{C}}^2\otimes {\mathbb{C}}^2$ is 3 it can always be spanned by product states. To see that this is the case, consider the subspace orthogonal to the most general state of two qubits\fnm{a}\fnt{a}{From now on we consider that the local bases $\{\ket{0},\ket{1}\}$ are not fixed for any party and therefore they include any local unitary transformation.} $\ket{\psi^\bot}=\alpha\ket{00}+\beta\ket{11}$, with $\alpha^2+\beta^2=1$. This subspace can be spanned by the states $\{\ket{10},\ket{01},(\beta\ket{0}-\alpha\ket{1})(\ket{0}+\ket{1})/\sqrt{2}\}$, which are all product states.

The remaining case is when the dimension of the subspace is 2. There are two possibilities \cite{sanpera98}, either the subspace contains only one product state and it is spanned by $\{\ket{00},(\alpha\ket{01}+\beta\ket{10}+\gamma\ket{00})\}$, with $\alpha^2+\beta^2+\gamma^2=1$, or it is spanned by the product states $\{\ket{00},(\kappa\ket{0}-\lambda\ket{1})(\mu\ket{0}+\nu\ket{1})\}$, with $\kappa^2+\lambda^2=1$ and $\mu^2+\nu^2=1$.

Therefore, the only case that satisfies the necessary condition is when the rank of both states of two qubits is 2 and their ranges contain only one product state. In fact, since a mixed state of two qubits of rank two can always be regarded as a pure state of three qubits where one of the parties has been traced out, we can use the known results of the classifications of pure states in ${\mathbb{C}}^2\otimes {\mathbb{C}}^2\otimes {\mathbb{C}}^2$ \cite{dur00,acin00}. The case in which the range only contains one product state corresponds to states of three qubits belonging to the W class \cite{dur00}. Since the states belonging to this family can always be written as $\ket{\Phi_W}_{ABC}= \sqrt{p}\ket{\Phi}_{AB}\ket{1}_C+\sqrt{1-p}\ket{00}_{AB}\ket{0}_C$, with $\ket{\Phi}_{AB}=\alpha\ket{01}+\beta\ket{10}+\gamma\ket{00}$ and $\alpha^2+\beta^2+\gamma^2=1$, we only have to consider the two-qubit mixed states of the form
\be\label{state2}
\rho_{AB}=p\proj{\Phi}+(1-p)\proj{00}
\ee

In the following, we will show how an entangled pure state can be purified from two states of two qubits of the form (\ref{state2}).

Let us consider the states of two qubits
\bea\label{states}
\rho_{AB}&=&p\proj{\varphi}+(1-p)\proj{00}\nonumber\\
\sigma_{A'B'}&=&q\proj{\psi}+(1-q)\proj{00},
\eea
where
\bea
\ket{\varphi}_{AB}&=&\alpha\ket{01}+\beta\ket{10}+\gamma\ket{00}\nonumber\\
\ket{\psi}_{A'B'}&=&\alpha'\ket{01}+\beta'\ket{10}+\gamma'\ket{00},
\eea
with $\alpha^2+\beta^2+\gamma^2=1$ and $\alpha'^2+\beta'^2+\gamma'^2=1$ (we will put the sub-indices denoting the parties whenever it is not clear from the context). One of the parties will hold the subsystem $AA'$ and the other the subsystem $BB'$.

The protocol will consist in general of two operators acting locally on each side, $M_{AA'}$ and $N_{BB'}$. There are some constraints on the operator $M_{AA'}\otimes N_{BB'}$ due to the fact that its kernel must contain all the product states (with respect to $AA'$ and $BB'$) in $R(\rho_{AB}\otimes\sigma_{A'B'})$. The reason is that there is no way of obtaining an entangled state from a product one, and therefore if there is a product state $\ket{e}_{AA'}\otimes\ket{f}_{BB'}$ in the range of the mixed state $\rho_{AB}\otimes\sigma_{A'B'}$ it has to be necessarily projected away by $M_{AA'}\otimes N_{BB'}$ (remember the protocol works for any state with the same range as $\rho\otimes\sigma$), i.e.
\be
M_{AA'}\otimes N_{BB'}\ket{e}_{AA'}\otimes\ket{f}_{BB'}=0, \hspace{.5cm}\forall \;\ket{e}_{AA'}\otimes\ket{f}_{BB'}\in R(\rho_{AB}\otimes\sigma_{A'B'}).
\ee

The range of the state $\rho_{AB}\otimes\sigma_{A'B'}$ is spanned by the states
\bea
\ket{00}_{AB}&\otimes&\ket{00}_{A'B'}\nonumber\\\ket{00}_{AB}&\otimes&\ket{\psi}_{A'B'}\nonumber\\ \ket{\varphi}_{AB}&\otimes&\ket{00}_{A'B'}\nonumber\\\ket{\varphi}_{AB}&\otimes&\ket{\psi}_{A'B'}.
\eea
Using the constrain for the operator $M_{AA'}\otimes N_{BB'}$ stated above we can see that the following relation has to be satisfied (from now on we will use a shorthand notation without the tensor product symbol unless it is misleading)
\be\label{constrain1}
M_{AA'} N_{BB'}\ket{00}_{AB}\ket{00}_{A'B'}=0,
\ee
which means that either (i) $M_{AA'}\ket{00}_{AA'}=0$ or (ii) $N_{BB'}\ket{00}_{BB'}=0$ have to be satisfied. We can choose (ii) without loss of generality and apply this condition to the states $\ket{\varphi}_{AB}\ket{00}_{A'B'}$ and $\ket{00}_{AB}\ket{\psi}_{A'B'}$
\bea\label{constrain2}
M_{AA'} N_{BB'}\ket{\varphi}_{AB}\ket{00}_{A'B'}\propto M_{AA'} N_{BB'}\ket{00}_{AA'}\ket{10}_{BB'}\nonumber\\
M_{AA'} N_{BB'}\ket{00}_{AB}\ket{\psi}_{A'B'}\propto M_{AA'}
N_{BB'} \ket{00}_{AA'}\ket{01}_{BB'}.
\eea
From (\ref{constrain2}) it is clear that a further constrain has to be imposed on the operators $M_{AA'}$ and $N_{BB'}$, as the states appearing in the r.h.s. of (\ref{constrain2}) are product states. Since $N_{BB'}$ cannot be orthogonal to both $\ket{01}_{BB'}$ and $\ket{10}_{BB'}$ (it would make $r(N_{BB'})=1$ and then the entanglement between $AA'$ and $BB'$ would be destroyed) we conclude that $M_{AA'}$ has to be orthogonal to $\ket{00}_{AA'}$, i.e. condition (i) stated above has to be satisfied as well.

Finally, we can apply both conditions ---$(i)$ and $(ii)$--- to the state $\ket{\varphi}_{AB}\ket{\psi}_{A'B'}$, obtaining
\be\label{constrain3}
M_{AA'} N_{BB'}\ket{\varphi}_{AB}\ket{\psi}_{A'B'}=
M_{AA'}N_{BB'}(\alpha'\beta\ket{10}_{AA'}\ket{01}_{BB'}+\alpha\beta'\ket{01}_{AA'}\ket{10}_{BB'}).
\ee
Note that the terms that do not appear in the r.h.s. of (\ref{constrain3}) have been projected away, since they contained either $\ket{00}_{AA'}$ or $\ket{00}_{BB'}$ which are orthogonal to $M_{AA'}$ and $N_{BB'}$.

Now we only have to apply the optimal protocol to obtain a maximally entangled state from a pure state \cite{vidal99,lo01}, which gives
\bea\label{operation1}
M_{AA'}&=&\min\{\alpha\beta',\alpha'\beta\}\left(\frac{1}{\alpha\beta'}\proj{01}+ \frac{1}{\alpha'\beta}\proj{10}\right)\nonumber\\
N_{BB'}&=&\proj{01}+\proj{10},
\eea
and the probability of obtaining a maximally entangled state from
$\rho\otimes\sigma$ in (\ref{states}) is
\be\label{prob}
P(\rho\otimes\sigma\rightarrow\Psi)=2 p q
\min\{\alpha^2\beta'^2,\alpha'^2\beta^2\}
\ee

It is important to note that this is the \emph{optimal} probability, since we have obtained (\ref{constrain3}) only by applying the constrains imposed by the range of the mixed state $\rho\otimes\sigma$ and (\ref{operation1}) is the optimal protocol for the conversion between pure states. 

We can summarize the result we have obtained in the following way:

\vspace*{12pt}
\noindent
{\bf Result 1:} Two mixed states of two qubits can be purified to a pure entangled state if and only if their ranges cannot be spanned by product states. In that case, the optimal probability is given by (\ref{prob}).


A further question is the condition under which a pure entangled state can be purified from n copies of a two-qubit mixed state. That is, when the transformation using local operations and classical communication
\be
\rho_{AB}^{\otimes n}\rightarrow \proj{\Psi},
\ee
is possible with $n$ finite and $\ket{\Psi}$ being an entangled state.

From the results we have obtained, and since in this case we are dealing with copies of the same state, we have that the necessary and sufficient condition stated above is still valid. 

\vspace*{12pt}
\noindent
{\bf Result 2:} $n$ (finite) copies of a two-qubit mixed state $\rho_{AB}$ can be purified to a pure entangled state if and only if $n\geq 2$ and the range of $\rho_{AB}$ cannot be spanned by product states.

\vspace*{12pt}
\noindent
{\bf Proof:} The proof is similar to that of the necessary condition. Any protocol purifying a pure entangled state from $n$ copies of $\rho_{AB}$ would work for any state $\sigma_{AB}$ with the same range as $\rho_{AB}$. This makes it impossible to distill a pure entangled state from a density matrix $\rho_{AB}$ whose range can be spanned by product states even when we have $n$ copies of it. On the other hand, from \emph{result 1} obtained before we know that the necessary condition is actually sufficient for two-qubit mixed states, and therefore the result is proven.

This result is interesting, since it means that only a very small subset of the (entangled) mixed two-qubit states can be purified to a pure entangled state. On the other hand, all the entangled states of two qubits can be distilled, i.e. the Entanglement of Distillation is non-zero for all the entangled two-qubit states. The reason of this apparent contradiction is the following. As explained in the introduction, the distillation protocols are defined in the asymptotic limit. The protocol transforms the initial state $\rho_{AB}^{\otimes{n}}$ into the state $\Omega_{AB}$, that has a fidelity with the state $\proj{\Psi_-}^{\otimes n E_D(\rho_{AB})}$ that approaches 1 when the number of copies $n$ goes to infinity,
\be
\lim_{n\rightarrow\infty} \mathrm{Tr}\left[\Omega_{AB}, \proj{\Psi_-}^{\otimes n E_D(\rho_{AB})}\right]\longrightarrow 1,
\ee
where $E_D(\rho_{AB})$ is the entanglement of distillation of $\rho_{AB}$ and in this case the fidelity is the trace. Our results only apply for the case of finite $n$, and in this case the distillation protocols are approximate transformations giving in general a mixed state as a result. 


Summarizing, in this paper we have found the optimal protocol to purify two mixed states of two qubits. The result reveals that only a very small subset of the mixed two-qubit states can be purified. Moreover, we have given a simple necessary and sufficient condition to determine whether a mixed two-qubit state can be purified when a finite number of copies are available. 

\nonumsection{Acknowledgements}

The author thanks R. Tarrach for posing this problem and for many interesting discussions and useful comments on the final manuscript. The author also thanks A. Ac\'{\i}n, L. Masanes and G. Vidal for helpful comments and suggestions. This work is financially supported by MEC (AP99), AEN99-0766, 1999SGR-00097 and IST-1999-11053. 

\nonumsection{References}

\end{document}